\documentclass[twocolumn]{aastex63}
\usepackage{natbib}
\usepackage{needspace}
\usepackage{tabularx}
\usepackage{booktabs}
\usepackage{amsmath}
\usepackage{tabularx}

\newcommand{\Tmag}{${T_{\mathrm{mag}}}$}

\graphicspath{{./}{figures/}}

\newcommand\totknowndips{500}
\newcommand\totcat{414}
\newcommand\totdippers{293}
\newcommand\newdippers{234}

\shortauthors{Capistrant et al.}
\begin{document}
\title{A Population of Dipper Stars from the Transiting Exoplanet Survey Satellite Mission}
\author[0000-0002-4592-8799]{Benjamin K.~Capistrant}
\altaffiliation{bcapistrant@ufl.edu}
\affiliation{Department of Astronomy,  University of Wisconsin-Madison, 475 N.~Charter St., Madison, WI 53703, USA}
\affiliation{Department of Astronomy, University of Florida, Gainesville, FL 32611, USA}

\author[0000-0001-7493-7419]{Melinda Soares-Furtado}
\altaffiliation{NASA Hubble Postdoctoral Fellow}
\affiliation{Department of Astronomy,  University of Wisconsin-Madison, 475 N.~Charter St., Madison, WI 53703, USA}

\author[0000-0001-7246-5438]{Andrew Vanderburg}
\affiliation{Department of Astronomy,  University of Wisconsin-Madison, 475 N.~Charter St., Madison, WI 53703, USA}
\affiliation{Department of Physics and Kavli Institute for Astrophysics and Space Research, Massachusetts Institute of Technology, Cambridge, MA 02139, USA}

\author[0000-0002-5365-1267]{Marina Kounkel}
\affil{Department of Physics and Astronomy, Vanderbilt University, VU Station 1807, Nashville, TN 37235, USA}

\author[0000-0003-3182-5569]{Saul A.~Rappaport}
\affiliation{Department of Physics and Kavli Institute for Astrophysics and Space Research, Massachusetts Institute of Technology, Cambridge, MA 02139, USA}

\author{Mark Omohundro}
\affiliation{Citizen Scientist, c/o Zooniverse, Department of Physics, University of Oxford, Denys Wilkinson Building, Keble Road, Oxford, OX1 3RH, UK}

\author[0000-0003-0501-2636]{Brian P. Powell}
\affiliation{NASA Goddard Space Flight Center, 8800 Greenbelt Road, Greenbelt, MD 20771, USA}

\author[0000-0002-5665-1879]{Robert Gagliano}
\affiliation{Amateur Astronomer, Glendale, Arizona, USA}

\author[0000-0003-3988-3245]{Thomas Jacobs}
\affiliation{Amateur Astronomer, 12812 SE 69th Place, Bellevue WA, USA}

\author[0000-0001-9786-1031]{Veselin~B.~Kostov}
\affiliation{NASA Goddard Space Flight Center, 8800 Greenbelt Road, Greenbelt, MD 20771, USA}
\affiliation{SETI Institute, 189 Bernardo Ave, Suite 200, Mountain View, CA 94043, USA}
\affiliation{GSFC Sellers Exoplanet Environments Collaboration}

\author[0000-0002-2607-138X]{Martti H.~Kristiansen}
\affiliation{Brorfelde Observatory, Observator Gyldenkernes Vej 7, DK-4340 T\o{}ll\o{}se, Denmark}

\author[0000-0002-8527-2114]{Daryll M.~LaCourse}
\affiliation{Amateur Astronomer, 7507 52nd Pl NE, Marysville, WA, 98270, USA}

\author[0000-0002-5034-0949]{Allan R.~Schmitt}
\affiliation{Citizen Scientist, 616 W. 53rd. St., Apt. 101, Minneapolis, MN 55419, USA}

\author[0000-0002-1637-2189]{Hans Martin Schwengeler}

\affiliation{Citizen Scientist, Planet Hunter, Bottmingen, Switzerland}

\author[0000-0002-0654-4442]{Ivan A.~Terentev}

\affiliation{Citizen Scientist, Planet Hunter, Petrozavodsk, Russia}

\begin{abstract}
Dipper stars are a classification of young stellar objects that exhibit dimming variability in their light curves, dropping in brightness by 10-50$\%$, likely induced by occultations due to circumstellar disk material.
This variability can be periodic, quasi-periodic, or aperiodic. 
Dipper stars have been discovered in young stellar associations via ground-based and space-based photometric surveys.
We present the detection and characterization of the largest collection of dipper stars to date:
\totdippers{} dipper stars, including \newdippers{} new dipper candidates.
We have produced a catalog of these targets, which also includes young stellar variables that exhibit predominately bursting-like variability and symmetric variability (equal parts bursting and dipping). 
The total number of catalog sources is \totcat{}.
These variable sources were found in a visual survey of TESS light curves, where dipping-like variability was observed.
We found a typical age among our dipper sources of $<$5\,Myr, with the age distribution peaking at $\approx 2$\,Myr, and a tail of the distribution extending to ages older than 20\,Myr. Regardless of the age, our dipper candidates tend to exhibit infrared excess, which is indicative of the presence of disks.
TESS is now observing the ecliptic plane, which is rich in young stellar associations, so we anticipate many more discoveries in the TESS dataset.
A larger sample of dipper stars would enhance the census statistics of light curve morphologies and dipper ages.
\end{abstract}


\section{Introduction}
\label{sec:intro}
The variability of young stellar objects (YSO) is extremely diverse, owing to a wide range of physical processes that are germane to early phases of stellar evolution. 
This includes chromospheric activity, the active accretion from circumstellar material that may generate hot spots and bursting activity, and occultations from a dusty circumstellar disk \citep{Cody_2014}.
Among the classes of YSO variables are the \textit{dipper stars}, a subclass of T Tauri stars that exhibit episodic or quasi-periodic dimming events where the brightness drops by ${\sim}$10-50$\%$ \citep[e.g.,][]{Alencar2010,Cody2010,Morales2011,Cody_2014,Stauffer2015,Ansdell2016,Rodriguez2017,Cody2018,Hedges2018,Bredall2020,Venuti2021,Roggero2021}.

The vast majority of cataloged dipper stars are K and M dwarfs and are therefore low in mass.
The dimming events generally occur over short timescales, ranging from hours to days, with obvious changes in both the shape and depth of the dips. 
These sources have also been shown to exhibit bursting-like signatures in addition to dips in brightness \citep[e.g.,][]{Cody_2014,Hedges2018,Bredall2020}.

Dipper stars have been detected over the past two decades via ground-based and space-based photometric instruments and surveys (optical and infrared) including the Convection, Rotation and Planetary Transits satellite (CoRoT) \citep{Baglin2006, Auvergne2009}, the Spitzer Infrared Array Camera \citep{Fazio2004, Werner2004}, KELT \citep{Pepper2007}, Kepler/K2 \citep{Borucki2010,Howell2014}, TESS \citep{Ricker2015}, the All-Sky Automated Survey for Supernovae (ASAS-SN) \citep{Kochanek2017}, and The Next Generation Transit Survey (NGTS)  \citep{Wheatley2018}.

One possible mechanism driving the dipping phenomena is the presence of dust occultations from the inner regions of a nearly edge-on circumstellar disk \citep[e.g.,][]{Stauffer2015,Bodman2017}.
It has also been noted that some dippers are capable of exhibiting periodic features at the stellar rotation frequency if the disk truncation radius is near the corotation radius \citep[]{Bouvier2007,Alencar2018}. The presence of a nearly edge-on circumstellar disk is supported by spectroscopic and spectropolarimetric observations \citep[e.g.,][]{Bouvier2003,Alencar2010}.
We refer the reader to \cite{Roggero2021} for an up-to-date and detailed discussion of the mechanisms driving the dipper phenomena.

Over \totknowndips{} dipper candidates have been cataloged to date.
\cite{moulton2020search} provides an in-depth review of dipper star detections and their corresponding catalogs. Dipper stars are generally members of young associations and have been detected within a number of young co-moving systems, including $\rho$ Ophiuchus (1\,Myr) \citep[e.g.,][]{Ansdell2016}; the Orion Nebula Cluster (1-3\,Myr) \citep[e.g.,][Moulton et al.~\textit{submitted}]{Morales2011, tajiri2020, moulton2020search}; the Lagoon nebula (1-3\,Myr); the Lupus region (1-3\,Myr) \citep[e.g.,][]{Bredall2020,Nardiello2020}; the Taurus association (3\,Myr) \citep[e.g.,][]{Rodriguez2017, Rebull2020, Roggero2021}; the open cluster NGC~2264 (3\,Myr) \citep[e.g.,][]{Alencar2010,Cody_2014};
$\gamma$~Velorum (3-4\,Myr) \citep{Nardiello2020}; Chamaeleon I and II (both 3\,Myr) \citep[e.g.,][]{Frasca2020,Nardiello2020}; Upper Scorpius (5-10\,Myr)  \citep[e.g.,][]{Ansdell2016, Cody2018, tajiri2020}; and Corona Australis association ($<$10\,Myr) \citep{Nardiello2020}.

The population of dipper stars in NGC~2264 comprise 20-30$\%$ of all classical T Tauri cluster members \citep{Alencar2010, Cody_2014}, as well as 30\% of disk-bearing stars in the Taurus cluster \citep{Roggero2021}.
The mechanism driving this dipping phenomena in NGC~2264 was investigated by \cite{Stauffer2015}, where they determined the dimming to be induced by disk occultations at/near the co-rotation radius.

The detection of more candidates is necessary to better understand the relationship between light curve morphologies, stellar characteristics (mass, age, etc.), and circumstellar disk characteristics.
In addition, dipper stars have much to offer low-mass, pre-main-sequence stellar models, including the role of flares \citep[e.g.,][]{Favata2005}, the shape and mass distribution within the inner circumstellar disk \citep[e.g.,][]{Bouvier1999}, and the evolution of the circumstellar and protoplanetary disks \citep[e.g.,][]{Bodman2017}. 
We refer the reader to \cite{Bodman2017} for a detailed discussion of the stellar constraints and diagnostics provided by dipper stars.

To expand the population of dipper catalog sources, we searched through millions of TESS light curves to identify and characterize dipper star candidates.
In this work, we report the discovery of \newdippers{} new dipper stars and examine their age and membership demographics.
In Section~\ref{sec:Data}, we describe our variable star dataset and illustrate the light curves of several targets. 
In Section~\ref{sec:Methods}, we present our vetting and classification methods, including our methods to quantify light curve morphology and provide stellar age estimations. In Section~\ref{sec:Results}, we present the results of our calculations.
In Section~\ref{sec:Discussion}, we summarize our findings and discuss prospects to further expand the dipper candidate population.

\section{Data}
\label{sec:Data}
\subsection{Pre-made Light Curves of Targets}
We searched for compelling dipper stars by visual inspection of millions of TESS light curves. The light curves used in this study come from a number of different sources, which are described below. 
These include the Cluster Difference Imaging Photometric Survey (CDIPS, 
\citealt{CDIPS}), MIT Quick Look Pipeline (QLP, \citealt{QLP, QLP2}), PSF-based Approach to TESS High quality data Of Stellar clusters (PATHOS, \citealt{PATHOS}), the NASA Goddard Space Flight Center Eclipsing Binary survey (GSFC EB, \citealt{gsfclightcurves}), difference-imaging extracted light curves from \citet{Oelkers2019}, and the Science Processing Operations Center (SPOC, \citealt{SPOC}). 
Each of these databases makes use of TESS full frame images (FFIs), except for SPOC which uses 2-minute and 20-second Target Pixel Files (TPFs).

\begin{itemize}
    \item [--] \textbf{CDIPS:} The CDIPS team creates light curves from TESS full frame images with 30-minute cadences using difference imaging (also commonly referred to as ``image subtraction"). They target stars in open clusters, moving groups, as well as stars that show evidence of youth. 
    The ``cdips-pipeline" provides light curves for stars brighter than a Gaia magnitude of 16 with 20--25 day observation windows \citep{cdipspipeline}. 
    \item[--] \textbf{QLP:} The MIT QLP team uses the MIT quick look pipeline to create light curves from TESS FFIs for stars brighter than a TESS magnitude (\Tmag{}) of 13.5. 
    In support of the team's primary goal of generating light curves for promising exoplanet candidates, they provide single sector light curves for targets observed over the TESS Primary Mission, including sectors 1-26, and now the first year of the TESS Extended Mission, sectors 27 through 39. 
    Targets from the second year of the TESS Extended Mission are to be released in 2022.
    \item[--] \textbf{PATHOS:} The PATHOS project uses empirical Point Spread Functions (PSFs) to obtain light curves from TESS FFIs by minimizing dilution effects in crowded environments. Their methodology also includes aperture photometry to compare with their PSF fitting methods. 
    Their typical targets are stars in the faint regime (\Tmag{}~$>$13\,mag), and they focus on clusters that include many young stars. 
    \item[--] \textbf{GSFC EB Survey:} Some of our team members from the NASA Goddard Space Flight Center (GSFC) Astrophysics Science Division are responsible for the production of TESS FFI light curves for stars \Tmag{}~$<15$\,mag (in this paper, we refer to this sample as ``GSFC"). 
    A parallelized implementation of the \texttt{eleanor} Python module is then performed on all sources \citep{Feinstein2019}, and a catalog of eclipsing binaries is generated using a neural network classifier.
    In addition to the identification of eclipsing binaries, some dipper candidates were culled from this effort. We refer the interested reader to Section~2 of \cite{Powell2021} to learn more about the GSFC light curve reduction and classification effort.
    
    \item[--] \textbf{Oelkers \& Stassun Sample:} \cite{Oelkers2019} publicly released a population of TESS FFI light curves, which were reduced using a difference imaging technique that is described in \cite{Oelkers2018}.
    The target sources were extracted from the TESS Input Catalog. 
    The reduced light curves achieve a noise floor of 60\,ppm\,hr$^{-1/2}$.
    These light curves are available on the Filtergraph data visualization platform.\footnote{\url{https://filtergraph.com/tess_ffi}}
    
    \item[--] \textbf{SPOC:} The Science Processing Operations Center (SPOC) at NASA Ames Research Center processes data to generate a variety of information on target stars, including the light curves investigated in this study. 
    The reduction pipeline processes raw pixels, extracting photometry and astrometry for each target and correcting for systematic errors. 
    The pipeline catalogs target stars and includes their corresponding characteristics and pixel data \citep{SPOC}. In contrast to the other pipelines, the SPOC pipeline uses stationary aperture photometry to generate light curves, focusing on individual target files with 2-minute or 20-second data rather than FFIs.
\end{itemize}

\subsection{Identification of Variable Star Candidates}
As part of the \textit{Visual Survey Group} (VSG) collaboration, several of us performed a visual inspection of millions of these light curves.
The VSG is responsible for a number of discoveries using Kepler, K2, and TESS data sets \citep[e.g.,][]{Schmitt2016,Mann2020,Powell2021,Kristiansen2022}. 

To identify dipper candidates (among other variable star classes), the VSG team uses a Windows-based publicly available software system known as \texttt{LcTools} \citep{schmitt2019lctools,schmitt2021lctools}. \texttt{LcTools} provides users with three principal tools: \texttt{LcGenerator} to create normalized light curves in bulk, \texttt{LcSignalFinder} to automatically find and record candidate signals (periodic and aperiodic) in large sets of light curves, and \texttt{LcViewer} to visually inspect light curves for candidate signals.
We identified a total of 477 dipper star candidates from our survey of these datasets. 

\section{Methods}\label{sec:Methods}
\subsection{Identifying Sources of Dipping Variability}
\label{sec:target}
Our dipper candidates come from several high-level science products with differing reduction techniques, making it critical to confirm light curve variability for all sources.
Because TESS has large pixels  (${\sim}20$\,arcseconds) blending from nearby sources is common, making it difficult to identify primary sources of variability.

We therefore performed our analysis to identify blended signals using the publicly available \texttt{Lightkurve} python package \citep{Lightkurve}. We constructed and inspected customized light curves for the dipper star candidates using TPF\footnote{\url{https://heasarc.gsfc.nasa.gov/docs/tess/Target-Pixel-File-Tutorial.html}} cutouts from the TESS full frame images, generated by the TESSCut software\footnote{\url{https://mast.stsci.edu/tesscut/}} \citep{Astrocut}. 
TESSCut returns pixel time series from regions of the TESS full frame images in the vicinity of a target star. These cut-out images allow for detailed inspection of the star's light in a given observation and the production of custom light curves. 
To generate these light curves, we employed customized apertures, which mitigate blending from neighboring stars and can be used to more carefully identify the target responsible for the variability signal.

To generate light curves for our candidates, we first selected the pixel where the target star is located, masking surrounding pixels to isolate the flux coming from the target. 
For pixel selection, we use the World Coordinate System (WCS), included in the metadata of the \texttt{TessTargetPixelFile} objects. 
The fits headers from the WCS provide information on how the pixels map to celestial coordinates, which we use to locate the target star within a given TPF. 

\begin{figure*}[!p]
\centering
\includegraphics[width=\textwidth]{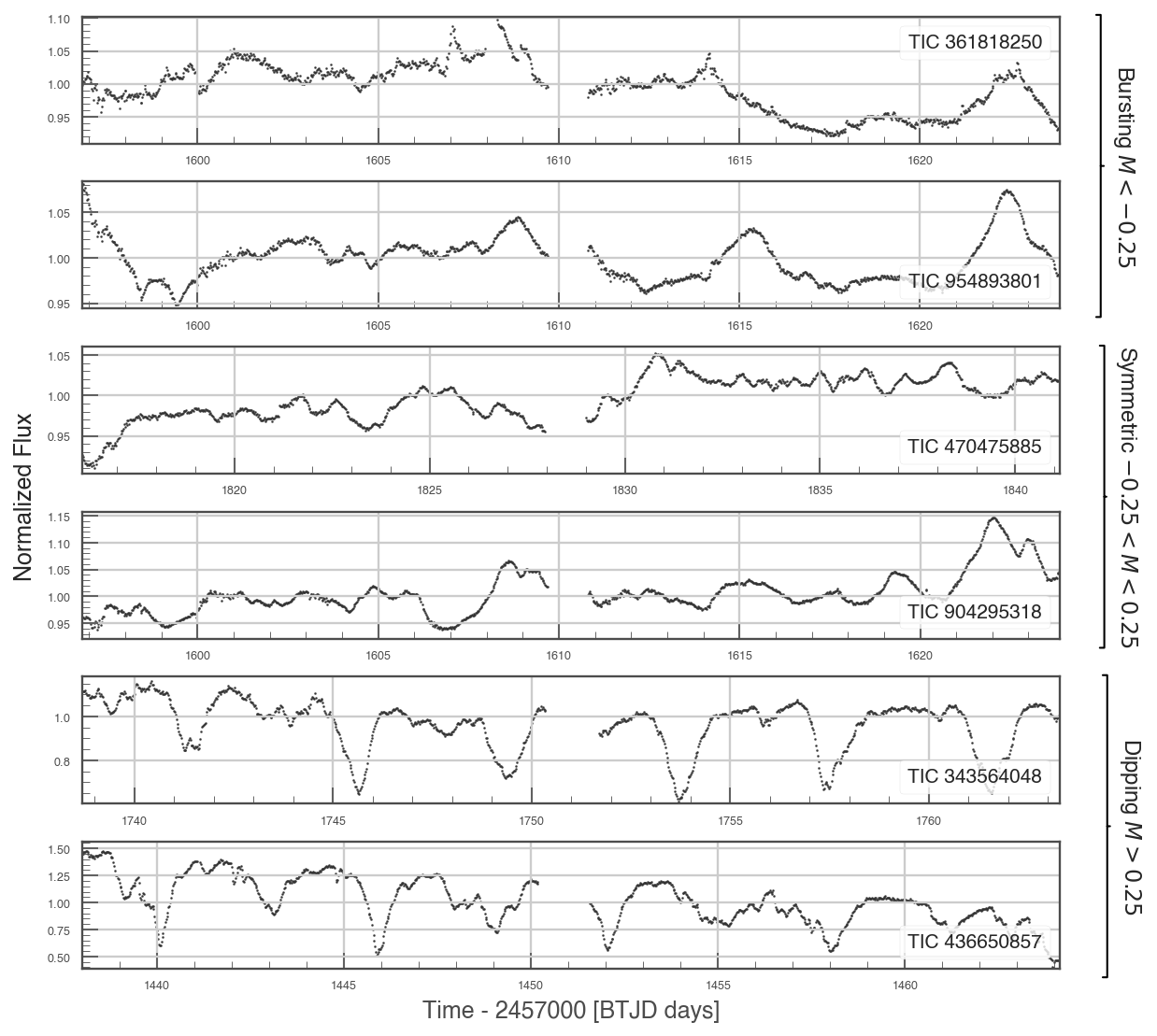}
\caption{Our custom light curves corresponding to six ``on target" dipper candidates. Included are TICs 361818250, 954893801, 470475885, 904295318, 343564048, and 436650857. 
Light curves were generated after isolating candidates in their TPFs. We opted to illustrate a range of flux asymmetry ``$M$" classifications as described in Section \ref{sec:M}. The top two light curves are examples of the predominately ``bursting" flux asymmetry ($M< -0.25$), rows three and four illustrate sources with ``symmetric" flux asymmetry ($-0.25<M<0.25$) and the bottom two rows illustrate sources  with predominately ``dipping" flux asymmetry ($M > 0.25$). Only sources with $M > 0.25$ are classified as dippers in our resulting catalog.}
\label{fig:DipLC}
\end{figure*}

We chose to customize our apertures for our targets because the \texttt{Lightkurve} package's built-in methods do not consistently mask the appropriate target star, especially when the target star is fainter than close-by companions. This is because the method relies on a set flux threshold, which does not always contain the target star. We chose the smallest possible aperture to confirm that the dipping variability is on or off-target. 
We identified the pixel closest to the location of the star, and then employed the \texttt{Lightkurve} function \texttt{tpf.to\_lightcurve()} to create a light curve from the isolated pixel, subtracting the light curve generated from the background pixels. 
Figure \ref{fig:DipLC} shows six light curves from our sample generated with our TPF masking method. 

\begin{figure*}[t!]
  \begin{center}
      \leavevmode
              \begin{minipage}[c]{0.49\textwidth}
\includegraphics[width=1.0\textwidth]{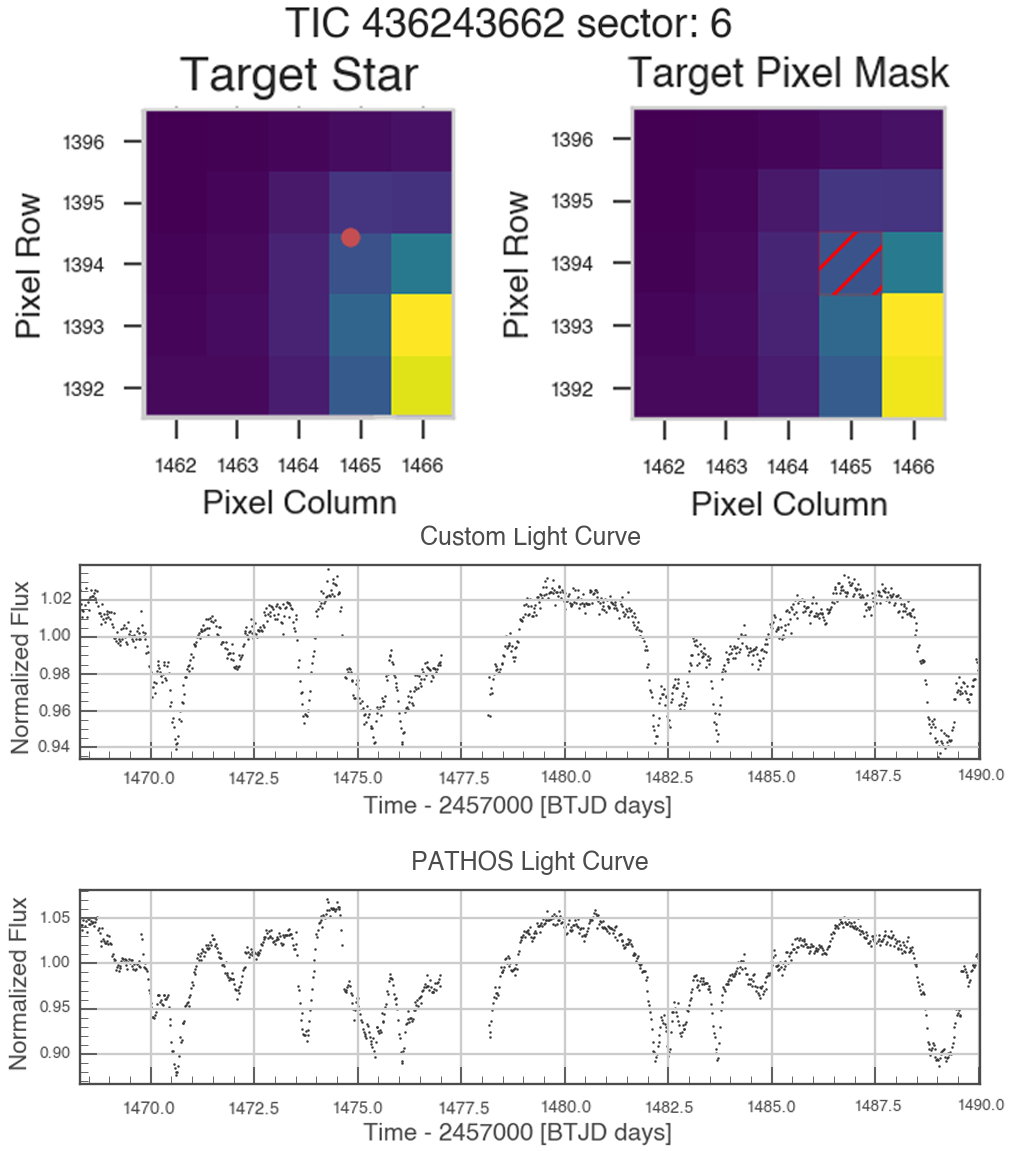} 
  \end{minipage}\hfill
  \begin{minipage}[c]{0.49\textwidth}
\includegraphics[width=1.0\textwidth]{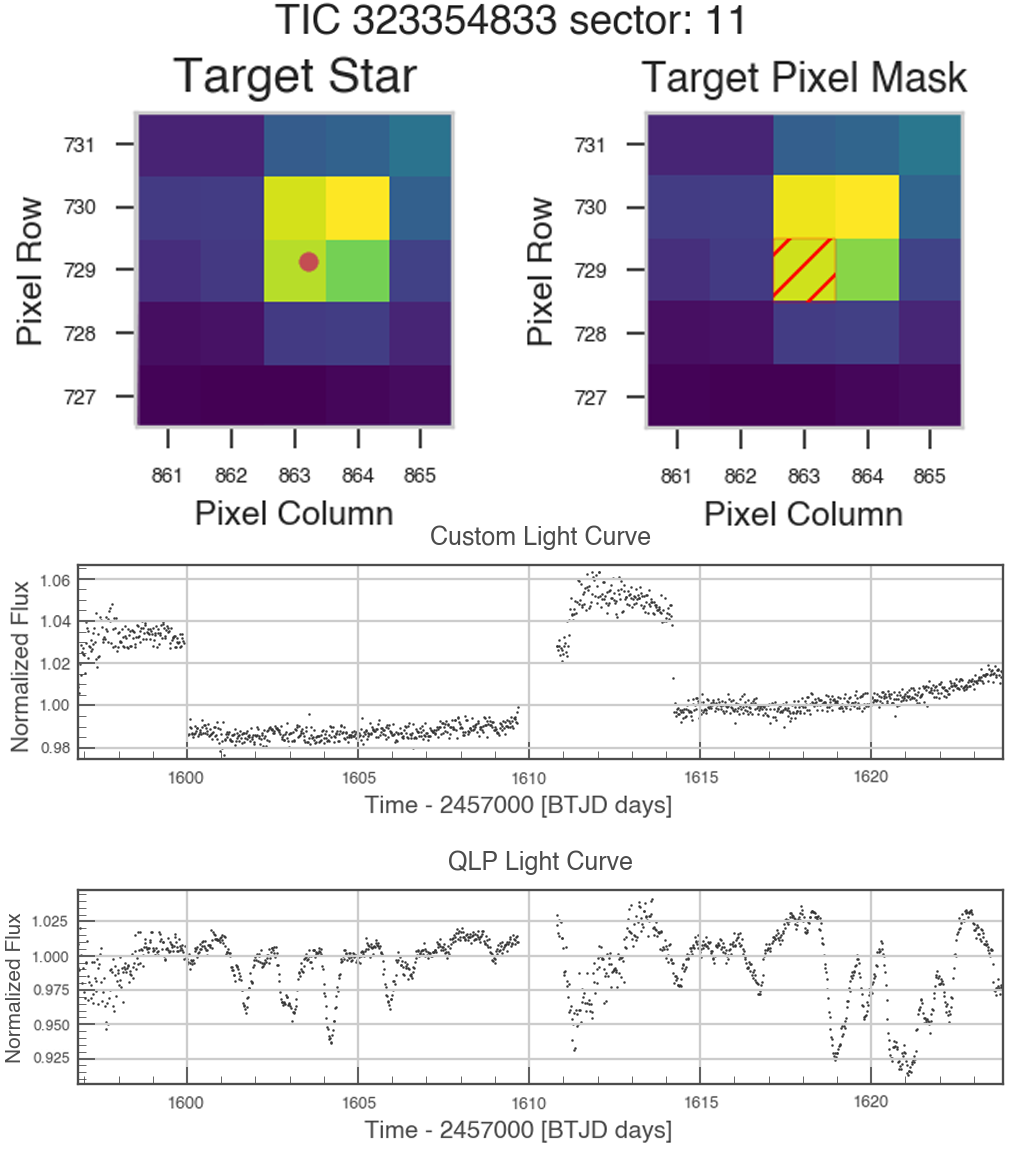} 
\label{allground}
  \end{minipage}
\end{center}
\caption{Example of an ``On Target" (left) and ``Off Target'' (right) sources. Top Row: Custom apertures applied to the TESS Target Pixel Files. Middle Row: Light curve generated from the masked pixel with background light curve subtracted. Bottom Row:
Pre-generated light curve (origin of the light curve varies depending on availability for each star). The left-hand example (TIC 436243662) shows similar variability in the pre-generated light curve and our custom light curve, consistent with the source of the variability being the target star. On the other hand, our custom single-pixel light curve in the right-hand example (TIC 323354833) does not show the same variability as the pre-generated light curve, indicating that the source of the variability may be off-target. } \label{onoffexample}
\end{figure*}

We classified stars as ``On Target" if the variability in our masked TPF light curves closely matched the variability in the pre-generated light curves. 
Similarly, sources were classified as ``Off Target" when the variability did not match. 
The On/Off Target determination was performed by eye, with each source visually inspected by three different individuals. 
We classified \totcat{} of the dipper candidates as on-target, and use these as our catalog going forward. 
An example of both on and off target stars can be found in Figure~\ref{onoffexample}. 

\subsection{Light Curve Flux Asymmetry Classification}
\label{sec:M}
Once we obtained the on-target star candidates, we classified their variability. 
We categorized the variable stars as bursting, symmetric, or dipping, which are descriptors defined by the flux asymmetry with respect to reflection along the flux axis of its light curve. 
To determine the light curve flux asymmetry of our target stars, we calculated the ``$M$" metric, which is described in more detail in \cite{Cody_2014}.
Following their methodology, we first subtracted a smoothed version of the light curve generated by a boxcar with a width of 10\,days. 
We then clipped 5$\sigma$ outliers from the boxcar smoothed light curve. 
After taking the mean of the top 10\% and bottom 10\% flux values from the residual light curve, we find the asymmetry metric, $M$, given by

\begin{equation}
M=\frac{\left(\left\langle d_{10\%}\right\rangle-d_{\mathrm{med}}\right)}{\sigma_{d}}
\end{equation}
\smallskip
 
 \noindent where $\left\langle d_{10 \%}\right\rangle$ is the top and bottom decile mean, $d_{\mathrm{med}}$ is the median of the long-term trend and outlier removed light curve flux, and $\sigma_d$ is the standard deviation.
 \cite{Cody_2014} found that the $M$ metric is only sensitive to timescales below 15--20\,days. 
 
 We remove long-term stellar variability and instrumental artifacts by smoothing our light curves with the \texttt{keplersplinev2}\footnote{\url{https://github.com/avanderburg/keplersplinev2}} tool \citep{Vanderburg2014}.
 We employed a spline knot spacing of 10\,days, and subtracted the best-fitting spline from the original light curve before computing the $M$ metric.
 While \texttt{keplersplinev2} was produced to work with the Kepler data set, this tool also works well with the TESS data set. Figure~\ref{fig:DipLC} displays the light curves of six ``on target" sources ranging in $M$ metric classifications. 
 The top two rows display sources that are predominately bursting ($M<-0.25$), the central two rows display sources that predominately symmetric ($-0.25>M>0.25$), and the bottom two rows display sources that are predominately dipping ($M>0.25$).
 Only sources with $M>0.25$ are classified as dippers in our resulting catalog.
  
\subsection{Periodicity of Dipper Candidates}
\label{sec:Q}
In this section, we describe the method for classifying the periodicity of our catalog sources by implementing a second variability metric presented in \cite{Cody_2014}: the periodicity metric, ``$Q$". 
Following their methods for comparative consistency,  we computed an autocorrelation function (ACF) on the magnitude of the light curve to home in on the period of the source.

Before computing, we interpolated the data to fill any gaps in the time series data. 
We then calculated and recorded the largest local ACF peak. To find the strongest ACF peak, we used the \texttt{signal.find\_peaks} function from the \texttt{Scipy} Python package \citep{scipy}, specifying the following three criteria: (1) adjacent ACF peaks must be separated by a minimum of 15 points, (2) the peak prominence (a measure of the vertical distance between the peak and the surrounding baseline of the signal) must exceed $0.05$, and (c) bypass the first identified ACF peak, as this feature is associated with short lags.
Our peak prominence setting of $0.05$ permits the identification of ACF peaks for all dipper candidates.
For a highly-aperiodic target, such as with white noise time series data, the resulting Q value is near unity, which indicates aperiodic light curve morphology.
Challenges corresponding to the Q metric, particularly as it pertains to aperiodic sources, are discussed further in Section~\ref{sec:Results}.
This topic is also the subject of investigation in \cite{Hillenbrand2022}.

To better determine the best period of a given source, we followed the ACF calculation with a periodogram search within $15\%$ of the value identified by the peak ACF.
Since our target light curves do not exhibit well-defined time variation, we measured the period of a given source using the Stellingwerf Phase Dispersion Minimization (PDM) module (\texttt{astrobase.periodbase.spdm}) from the \texttt{Astrobase} Python package \citep{Stellingwerf1978}.
The PDM method utilizes a brute-force approach to data folding. 
A large ensemble of periods is considered in the folding of the data.
The data are then binned and the binned variance and overall variance of each trial are computed. 
The ratio of the binned variance to overall variance is calculated for each investigated period. 
False periods have a ratio close to unity, while this ratio is small for true periods.
The PDM algorithm calculates the periodicity of time series data without constraining the search to a preconceived functional form. 
While the PDM technique is capable of revealing box-liked and sinusoidal variations, it is not limited to the detection of variations of this form.
Therefore, the PDM technique is robust at detecting periodic signatures like those observed by periodic dippers, semi-periodic dippers, bursters, and heartbeat stars.

The strongest identified period in the resulting PDM power spectrum was then used to generate a phase-folded light curve. 
We then smoothed the phase-folded light curve using a boxcar with a width that is 25\% that of the period, and subtracted this boxcar from the light curve after long-term trends and 5$\sigma$ outliers were removed.
For an illustrated example of the process described, we refer the interested reader to Figure~28 in \cite{Cody_2014}. 

The periodicity metric, ``$Q$", is given by 
\begin{equation}
Q=\frac{\left(\mathrm{RMS}_{\mathrm{resid}}^{2}-\sigma^{2}\right)}{\left(\mathrm{RMS}_{\mathrm{raw}}^{2}-\sigma^{2}\right)}
\end{equation}
\smallskip
    
 \noindent where $\mathrm{RMS}_{\mathrm{resid}}^{2}$ is the RMS value of the light curve after the phase-folded, boxcar-smoothed light curve has been subtracted and $\mathrm{RMS}_{\mathrm{raw}}^{2}$ is the RMS of the raw light curve before this subtraction. The parameter $\sigma$ denotes the uncertainties which we obtain using a different method than \cite{Cody_2014}. 
 To calculate this uncertainty, we employed a few different methods to determine typical relationships between TESS magnitude and photometric precision for each light curve source. For the CDIPS and QLP light curves, we referred to the figures plotting the photometric precision against the TESS magnitudes. 
 For the CDIPS light curves, we employed Figure~14 in \cite{CDIPS}; for QLP, Figure~1 in \cite{QLP}. For PATHOS, we downloaded the full set of light curves from TESS Sector 1, measured the scatter in each light curve by calculating the point to point scatter \citep{aigrain}, and produced our own plot of precision as a function of TESS magnitude. 
 
 We digitized each of these plots using an online plot digitizer application,\footnote{\url{https://apps.automeris.io/wpd/}} obtaining the relationship between \Tmag{} and precision. 
 Authors of the CDIPS, QLP, and PATHOS pipelines gave their photometric precision measurements over timescales of 1\,hour or 6\,hours and assumed that precision scaled with $1/ \sqrt{\text{exposure time}}$. We converted these precision measurements to 30-minute \textit{RMS}, and find the uncertainties, $\sigma$, of our candidate light curves based on their stars' \Tmag{}. For the SPOC light curves, we took the full set of light curves from Sector 1, binned the two-minute-cadence light curves to 30 minute cadences, and calculated the median value of a running standard deviation filter over the full light curve. We then calculated the median photometric precision in bins as a function of TESS magnitude. Once we had relationships between \Tmag{} and precision for each light curve source, we interpolated them to the TESS band magnitude of each stellar variable to estimate the photometric precision of each light curve. 

\subsection{Stellar Age for Variable Stars}
\label{sec:AgeMeth}
\begin{figure*}[tbh!]
    \centering
    \includegraphics[width=\textwidth]{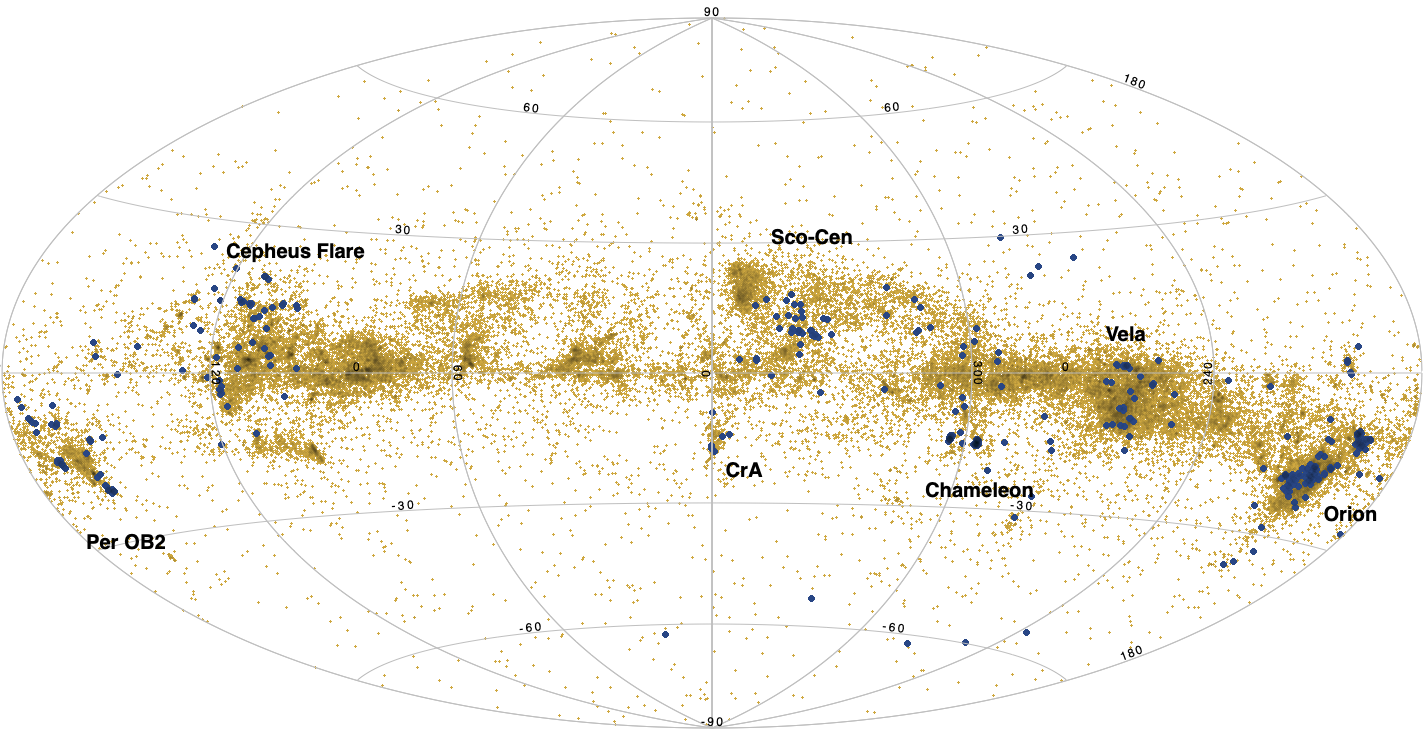}
    \caption{Spatial distribution of our catalog sources (blue) with several young stellar associations labeled. 
    Sources in young, co-moving groups are shown in gold.
    The distribution of our sample tracks with the location of these young stars.}
    \label{fig:map}
\end{figure*}

We compared the distribution of our catalog sources to known stellar populations. 
In Figure~\ref{fig:map}, we illustrate the spatial distribution of our catalog sources.
We found that approximately half of our variable stars were known members of young co-moving groups, as identified by \citet{kounkel2020}.
Most of the other catalog sources were found in known star-forming regions and may not have been included in the co-moving group catalog. 
Thus, to manually confirm their membership, we examined overdensities in the spatial distribution of the stellar variables in our sample towards star-forming regions such as Orion, Taurus, Sco Cen, Chameleon, Vela. 

We successfully implemented this method for a total of 371 out of \totcat{} stars (90\% of the sample).
The remaining 10\% of sources were either (a) found in the vicinity of very low mass star-forming regions, for which it is difficult to conclusively assign membership, (b) exhibit peculiar astrometry not fully consistent with the phase space of a given population, or (c) were found to be evolved stars that contaminated the sample.

Although average ages are available for each population of co-moving groups, those ages may not be accurate, due to the fact that an individual star-forming region can sustain star formation for $>$10\,Myr \citep[e.g.,][]{kounkel2018a,damiani2019}. 
As such, we opted to estimate the ages of the individual pre-main sequence stars using the neural network \texttt{Sagitta} \citep{mcbride2020}, which utilizes parallaxes, as well as Gaia and 2MASS photometry in G, GBP, GRP, J, H, and K bands. \texttt{Sagitta} has been trained on the photometry of young stars in populations with known ages, interpolating across a sample of $\sim$70,000 stars. 
This establishes an internal empirical age sequence, allowing for the minimization of systematic trends, such as those pertaining to stars of different masses having systematically different derived ages in the same cluster.
It also allows us to interpret the ages for the sources with significant reddening or infrared (IR) excess due to the presence of protoplanetary disks, which can be difficult to place on theoretical isochrones. The trained model can then be applied to the previously unseen data to estimate ages up to $\lesssim$40--70\,Myr.

\section{Results}\label{sec:Results}
From our initial sample of 477 light curves exhibiting variability, we determine \totcat{} candidates to be ``on target" sources, which are presented in our Variable Star Catalog. 
As described in Section~\ref{sec:target}, these stars were classified as ``on target" when the variability in our masked TPF light curves closely matched the pre-generated light curves.
We included only ``on target" candidates in our analyses of the morphology and ages of this population. 
In Table~\ref{tab:1}, we provide the first ten rows from our Variable Star Catalog.  
The listed properties were obtained using data from Gaia survey data (Data Releases 2 and EDR3), as well as the TESS input catalogs \citep[e.g.,][]{GaiaEDR3,GaiaDR2,TIC}.
Also included are our calculated ``$M$" and ``$Q$" variability metrics, the corresponding classification, the identification of new detections, and the estimated age of the target.
 
\subsection{Morphology}
Our catalog sources were initially flagged by the VSG as dipper candidates due to the visible dips in brightness observed in their light curve morphologies.
It is therefore not surprising that in our computation of the ``$M$" and ``$Q$" values, we classified the vast majority (70\%) of these sources as dipper stars. 
Figure~\ref{fig:DipLC} depicts example light curve morphologies for sources with differing values of ``$M$", including sources in the two top rows that are classified as bursters.
It is important to note that the sign of the flux asymmetry, ``$M$", value is inverted from the intuitive value based on flux, as the units are reflective of magnitude.
Therefore, bursting-like variability corresponds to a negative $M$ value, while dipping-like variability corresponds to a positive $M$ value.

\begin{figure*}[tbh!]
    \centering
    \includegraphics[width=0.8\textwidth]{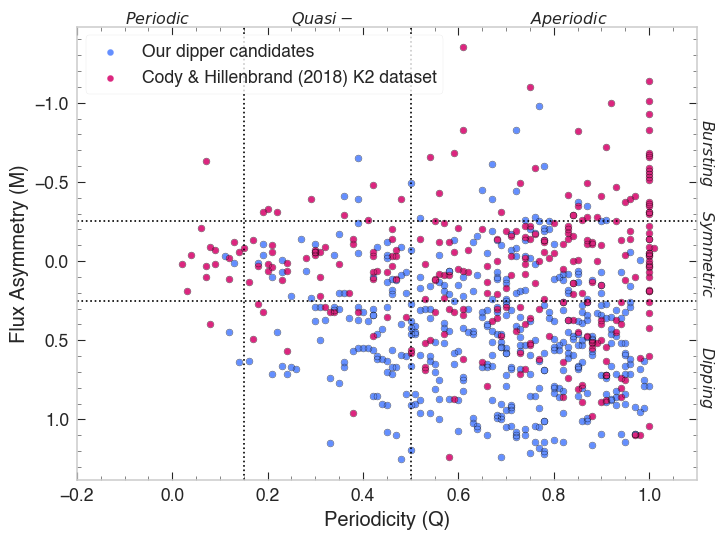}
    \caption{We plot the flux asymmetry ($M$) versus periodicity ($Q$) metrics for our stellar variables (blue), comparing these sources to the young stellar variables analyzed by \cite{Cody2018} (red). 
    Note that $M$ is reflective of magnitude, and the sign of the variable is inverted from the intuitive value based on flux. Therefore, dipper stars have positive rather than negative $M$ values. The majority of our catalog sources ($70\%$) exhibit predominately dipping flux asymmetries. This is expected based on the search criteria of our visual survey of TESS light curves.
    In accordance with \citet{Cody2018}, we find no clear correlation between $Q$ and $M$.}
    \label{fig:QM}
\end{figure*}

In Figure~\ref{fig:QM}, we illustrate the flux asymmetry versus periodicity of our variables. Bursters are shown in the top row, symmetric variables in the central row, and dipper stars in the bottom row. Stars classified with periodic variability are in the left column, quasi-periodic are in the central column, and those with aperiodic variability are in the right column.
This figure is akin to Figure~31 in \cite{Cody_2014}.
However, in this figure, we compare our sample to the population investigated by \cite{Cody2018} (red points) and refer the reader to their Figure~6, which illustrates the periodicity versus flux asymmetry of their sample. The targets from \cite{Cody2018} are members of the Upper Sco and rho Oph star-forming regions.
In their analysis, they made no selection for light curve morphology, and therefore their sample includes stars from all of the variability classes. 
The distribution of our catalog sources largely falls in the dipping region, which is in support of our expectation.
In agreement with \cite{Cody_2014}, we find no clear correlation between the $Q$ and $M$ statistics.

It bears mentioning that there are challenges associated with the variability metrics, especially with generating the $Q$ values. 
Firstly, identifying peaks in the ACF, an important step in determining the $Q$ metric, is not a straightforward process. 
In some cases, such as in the case of a highly-aperiodic light curve, this can impact the calculations. 
Another difficulty arises as a result of contaminants in the light curve. 
Contaminant may include rapidly rotating spotted stars; pulsating stars with varying amplitudes, phases, or periods; and hot exoplanets where the transit duration is shorter than the smoothing timescale.
Contaminants can result in the calculation of large $Q$ values. 
Despite these complications, the $Q$ metric is well-entrenched in scientific literature investigating YSO variability \citep[e.g.][]{Cody_2014,Cody2018,Rebull2015,Hillenbrand2022}.
We provide this metric for all our catalog variables to allow for comparison with previous and future work. 
Recently, \cite{Hillenbrand2022} determined that reduced cadence has a more pronounced effect on $Q$ than photometric uncertainty, and that $Q$ values become increasingly more inaccurate at larger, positive values.

\subsection{Age Distribution}
\label{sec:Ages}
\begin{figure}[tbh!]
    \includegraphics[width=0.46\textwidth]{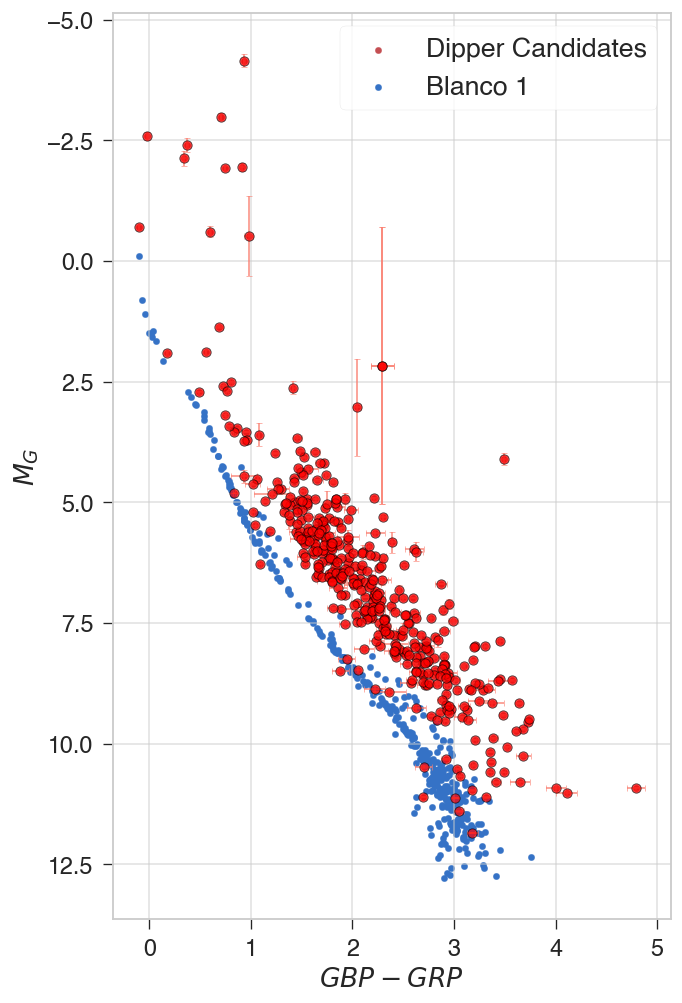}
    \caption{Diagram illustrating dipper candidates $GBP-GRP$ magnitudes against their absolute Gaia Magnitude from Gaia EDR3 catalog (415 sources shown as red points); Blanco 1 cluster (blue points) with an estimated age of 100-150\,Myr \citep{Cargile2010} are super-imposed for comparison. 
    With this figure, we can see our dipper candidates are overwhelmingly pre-main sequence sources.}
    \label{fig:HR}
\end{figure}
We illustrate the location of our dipper candidates in the HR diagram illustrated in Figure~\ref{fig:HR}.
To generate this figure, we employed data from the Gaia EDR3 catalog.
This figure includes a subset of our full catalog population, due to limits in the availability of Gaia EDR3 data for all sources.
As a point of comparison, we also show the HR diagram position of known members of the 100-150\,Myr old open cluster, Blanco~1, which is old enough that most of the stars (except very low-mass M-dwarfs) have reached their main-sequence HR diagram position. Most of our dipper candidates lie above the main sequence of the Blanco~1 population (blue points). Therefore, our sources are predominately pre-main sequence stars that are in the process of contracting onto the zero-age main sequence.

\begin{figure}[!t]
    \centering
    \includegraphics[width=0.5\textwidth]{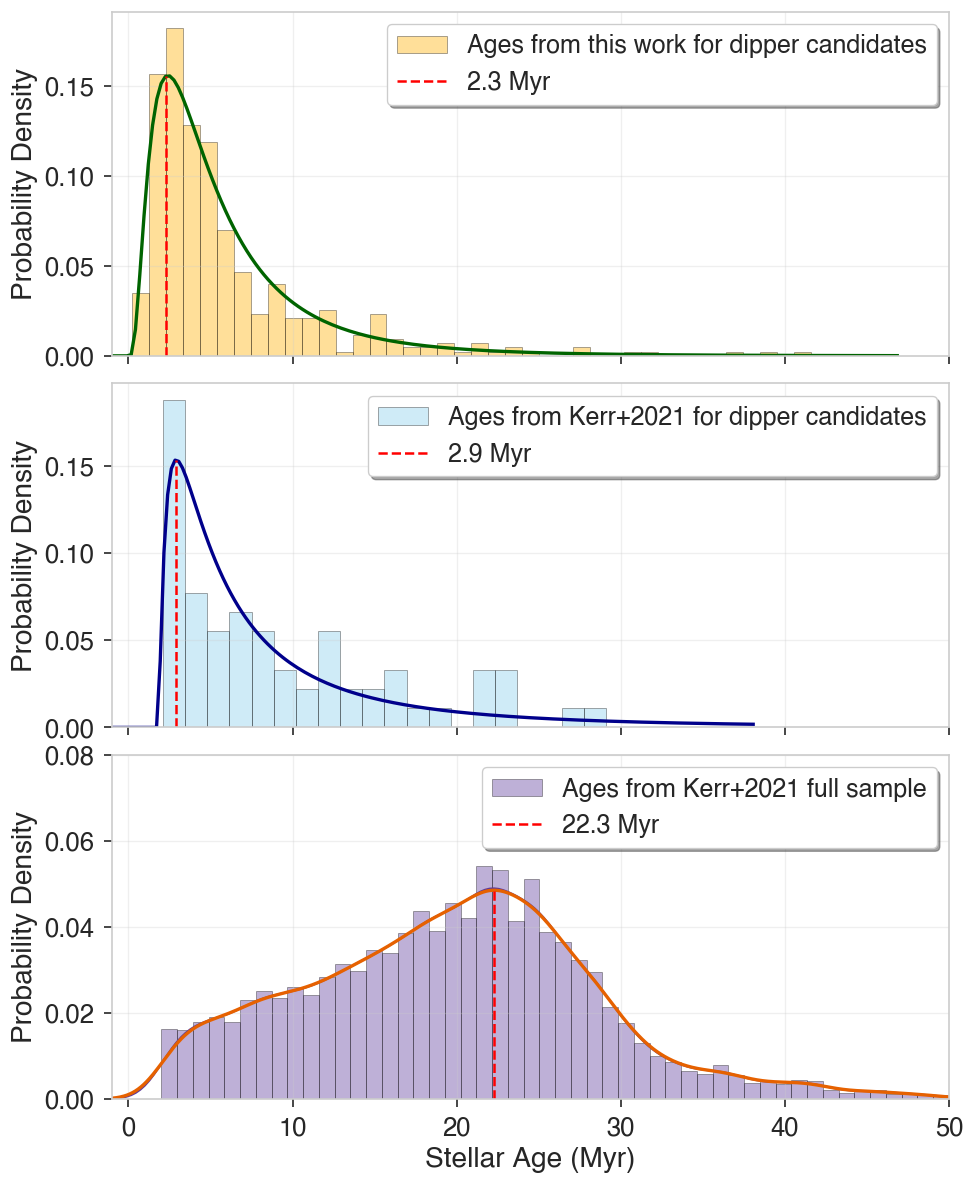}
    \caption{Top panel: Ages for all of our dipper candidates (\totcat{} stars), as determined by the \texttt{Sagitta} neural network \cite{mcbride2020}; Middle panel: Ages for a subsample of our dipper candidates (68 stars), as derived by \cite{Kerr2021}; Bottom panel: Ages for the entire sample (25,324 stars) in \cite{Kerr2021} for comparison. 
    In the top two panels, we fit the histograms with log-normal distributions to find the peak values, which we marked with a dashed red line. In the bottom panel, we used a kernel density estimate to determine the peak.}
    \label{fig:AgeComp}
\end{figure}

In the top panel of Figure~\ref{fig:AgeComp}, we illustrate the ages of our dipper candidates, as determined by the \texttt{Sagitta} neural network \citep{mcbride2020} (described in Section~\ref{sec:AgeMeth}). 
The dipper candidates in our sample exhibit a peak in their age distribution at 2.3\,Myr (red dashed line).
To determine the peak age of this distribution, we fit the histogram in the top panel with log-normal distribution. This method was also used to determine the peak age of the central panel.

\cite{Kerr2021} (hereafter K21) derived approximate ages for ${\sim}25,000$ Gaia DR2 catalog sources using photometric methods and a pipeline fully described in their study. Among the young stars in the K21 sample, 68 targets are dipper candidates found in our catalog.
In the central panel of Figure~\ref{fig:AgeComp}, we illustrate a histogram of the ages, as determined by the K21 analysis, of the overlapping K21 subsample of 68 targets. 
We observed that our dipper candidates are preferentially found among the youngest stars in their sample, with the age distribution peaking at 2.9\,Myr (red dashed line). 
In the bottom panel of Figure~\ref{fig:AgeComp}, we illustrate the age distribution of the full K21 sample, which exhibits a much broader range of ages, peaking at 22.3\,Myr. 
The peak in this distribution was determined using a kernel density estimate, as the full K21 population does not exhibit a log-normal distribution.

\begin{figure}
    \centering
    \includegraphics[width=0.5\textwidth]{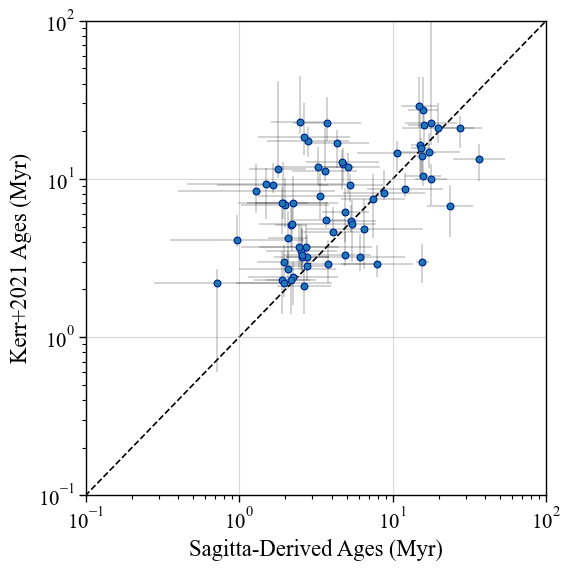}
    \caption{Scatter plot depicting the difference in the age calculations from the \texttt{Sagitta} neural network \citep{mcbride2020} and \cite{Kerr2021} sample for the overlapping subset of 68 stars found in both catalogs. The dashed diagonal shows what a one-to-one age relation would be. We find for a majority of the sources that ages derived by \cite{Kerr2021} are typically larger than those derived by \texttt{Sagitta}. The distribution of ages for these stars calculated by \cite{Kerr2021} can be seen in the middle panel of Figure \ref{fig:AgeComp}.}
    \label{fig:AgeDiff}
\end{figure}

We compare the ages of the overlapping 68 stars in the K21 sample and our catalog in Figure~\ref{fig:AgeDiff}. 
Overall, the distribution of ages between the two estimation methods is largely consistent within errors.
Yet, it bears mentioning that there are some systematic differences as well, largely induced by differences in the treatment of extinction in different regions, as well as due to different photometric bandpasses used (e.g., Gaia-only in \citet{Kerr2021} vs Gaia+2MASS in Sagitta). The outlying sources where the differences between the two approaches is most pronounced tend to be concentrated in a handful of specific regions, such as the Chameleon clouds.

\subsection{IR Excess}
\label{subsec:ir}
\begin{figure}[!t]
    \includegraphics[width=0.5\textwidth]{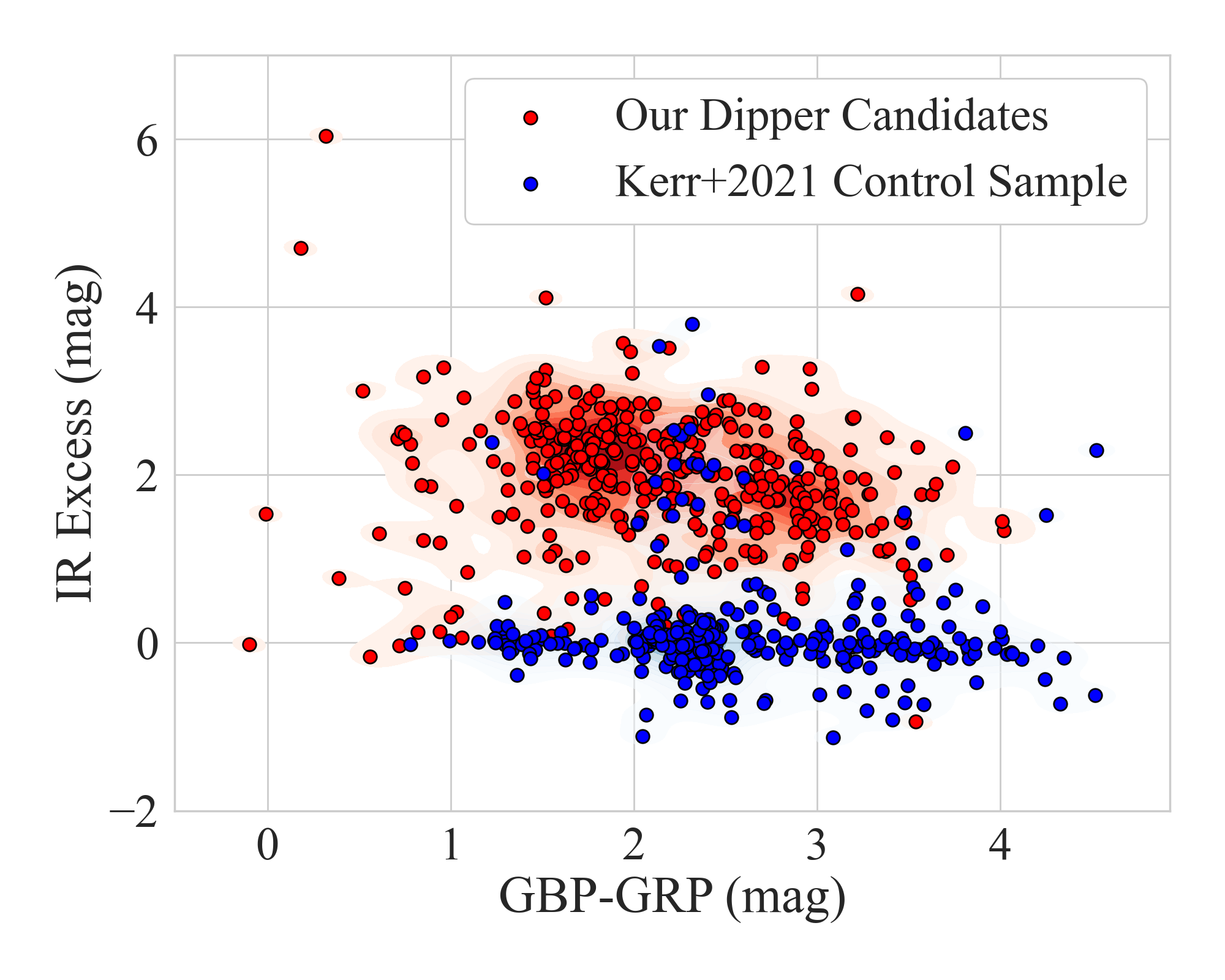}
    \caption{
    Scatter and contour plot illustrating the color ($GBP-GRP$) versus IR excess for our dipper candidates (red points), which range between 2-42\,Myr.
    A control sample of 392 sources from \cite{Kerr2021} (K21) are provided for comparison (blue points).
    A control star could be identified in the K21 sample for 93\% of our dipper candidates. 
    These comparison sources are close in age (within the K21 error bounds) to a given dipper candidate and have not been flagged as dipper variables in the past. 
    We see that across all $GBP-GRP$ values, our dipper candidates exhibit greater IR excess than the K21 control sample. 
    }
    \label{fig:irexcess}
\end{figure}
We also investigated our dipper candidates to determine if there were differences in the measurements of IR excess, as compared to young stars ($<50$\,Myr) that have not been flagged as candidate (or confirmed) dippers.
To determine IR excess, we combined data from \cite{Pecaut2013} and \cite{Kraus2014}, establishing an expected baseline $W1-W3$ color for young stellar sources ($<50$\,Myr). 
It was necessary to combine the two datasets to ensure full coverage of the $GBP-GRP$ range of our dipper candidates. 
The $GBP-GRP$ color information was obtained from the Gaia EDR3 catalog, while the IR measurements were taken from the WISE database \citep{wisecatalog}.
We then fit a second-order polynomial to the data, generating a function for the expected $W1-W3$ magnitude as a function of $GBP-GRP$. 
IR excess was calculated as the difference between a given dipper candidate's measured $W1-W3$ magnitude and the expected baseline $W1-W3$ value at the corresponding $GBP-GRP$ color. 

Our results are shown in Figure~\ref{fig:irexcess}, where we plot the IR excess of our dipper candidates as a function of their corresponding $GBP-GRP$ color (red points). 
As a comparison, we created a control sample (blue points) of young stars from \cite{Kerr2021}.
To generate this control sample, we performed a source-by-source age match for each dipper candidate.
More specifically, the control sample consists of stars that (1) have not been flagged as dipper variables in the past and (2) share a common age with a given dipper candidate (the closest age match possible within the error bounds of the K21 sample).
Using these constraints, we were able to identify a corresponding K21 control star for 93\% of our dipper candidates (N=392).

We see that across all $GBP-GRP$ values, on average, our dipper candidates exhibit greater IR excess than the K21 control sample. 
Fitting second order polynomials to the two populations in Figure~\ref{fig:irexcess}, we calculated the difference in IR excess between our dipper candidates to that of the K21 control sample. 
Among sources where $GBP-GRP=1$, the IR excess of our dipper candidates is 1.8\,mag greater than that of the K21 sample.
Similarly, among sources where $GBP-GRP=3$ the IR excess of our dipper candidates is 1.6\,mag greater than that of the K21 sample.
The targets in the K21 control sample likely owe their optical redness to the fact that they are embedded in dust, which has a direct impact on optical color but not IR color.
Our population likely owes their IR excess to the presence of disks. 

\subsection{Morphology and other parameters as a Function of Age}
\begin{figure*}[tbh!]
    \centering
    \includegraphics[width=\textwidth]{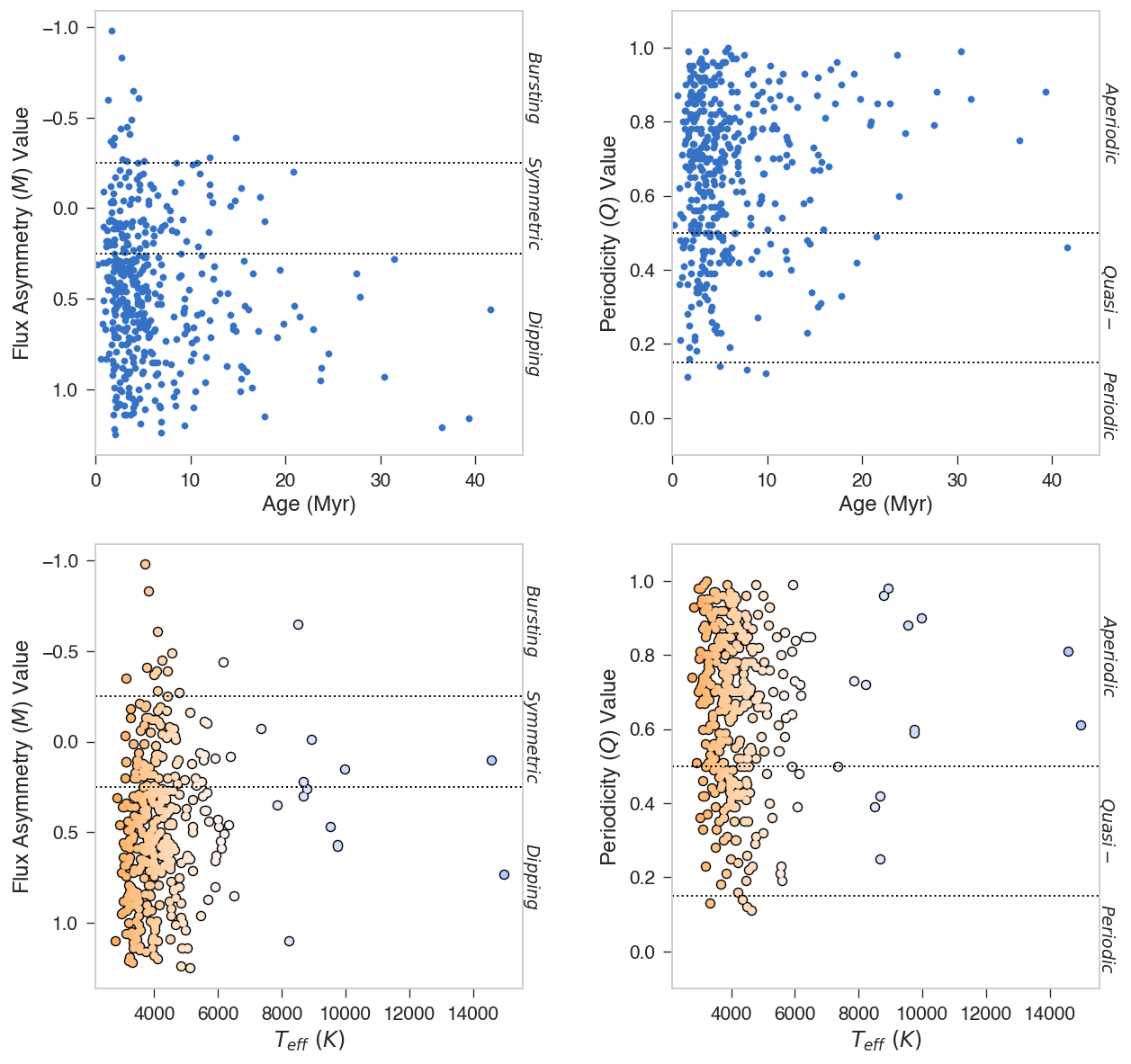}
    \caption{Stellar ages (Myr) determined in this work by the \texttt{Sagitta} neural network (as described by \citealt[][]{mcbride2020}) plotted against the calculated flux asymmetry metric, $M$ (upper left), and against the calculated periodicity metric, $Q$ (upper right). 
    In the bottom panels, we plot $T_{\mathrm{eff}}$ against the $M$ and $Q$ metrics. The points are the true color a human eye would see if viewing a blackbody of that temperature (pixel rgb values are obtained from \url{http://www.vendian.org/mncharity/dir3/blackbody/}). We find no clear correlations between age or temperature of the source and the variability metrics of the catalog sources.}
    \label{fig:AgeTQM}
\end{figure*}
We investigated for correlations between both the age and light curve morphology of our dipper candidates, as well as correlations between temperature and light curve morphology.
In the top row of Figure~\ref{fig:AgeTQM}, we observe a stochastic relationship between stellar age (in Myr) and both the flux asymmetry (left panel) and periodicity (right panel) metrics.
Similarly, in the bottom row of Figure~\ref{fig:AgeTQM}, we observe a stochastic relationship between the effective temperature and both the flux asymmetry (left panel) and periodicity (right panel) metrics.
In the bottom row, the points are color-coded to reflect the true color a human eye would see if observing a blackbody at that temperature.\footnote{\url{http://www.vendian.org/mncharity/dir3/blackbody/}}

\begin{figure*}[tbh!]
    \centering
    \includegraphics[width=\textwidth]{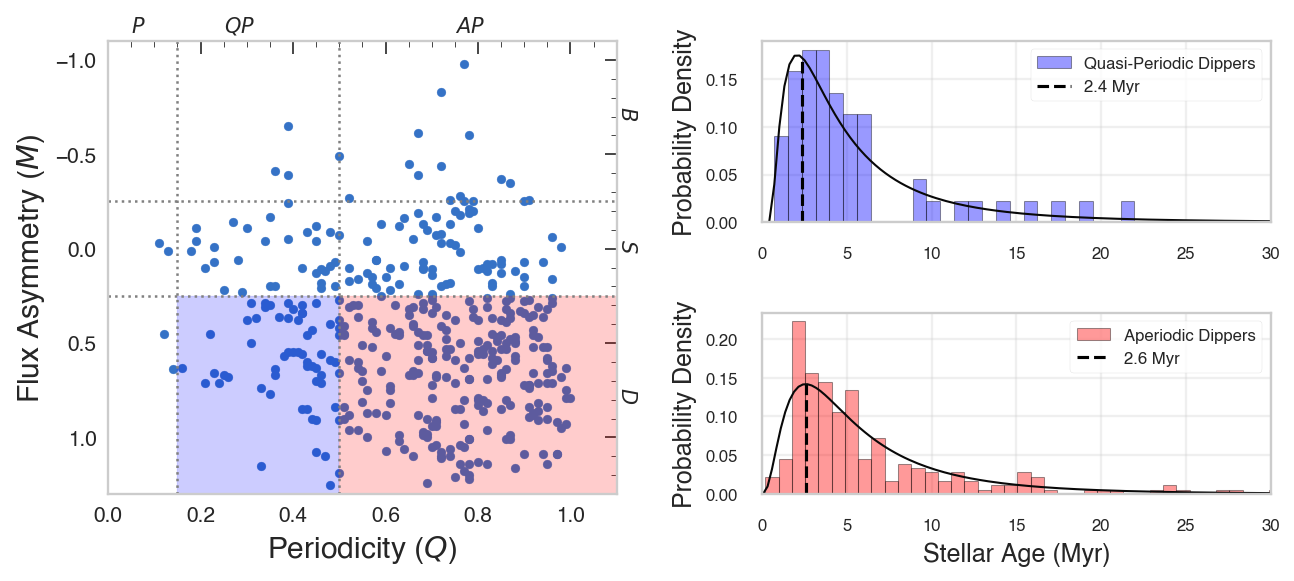}
    \caption{Sources from our catalog with ages determined by the \texttt{Sagitta} neural network \citep{mcbride2020}.
    Left: We split the dipper candidates by their flux asymmetries and periodicity metrics. Quasi-periodic dipper stars are shown in the blue region, and aperiodic dipper stars are shown in the red region.
    Right: Histograms of the ages calculated for the two groups of dipper stars, fit with log-normal distributions to identify the peak ages. 
    The distribution of ages for quasi-periodic  dipper stars peaks at 2.4\,Myr.
    Similarly, the age distribution of aperiodic dipper stars peaks at 2.6\,Myr. 
    As we found in Figure \ref{fig:AgeTQM}, these histograms show no indication of a relation between periodicity values and age for our dipper stars.}
    \label{fig:MedAge}
\end{figure*}

We also compared the age distributions between the quasi-periodic and aperiodic subgroups for our catalog dipper stars ($M>0.25$).
The two age distributions are illustrated in the right column of Figure~\ref{fig:MedAge}. 
Both distributions peak near a similar age, and a Kolmogorov-Smirnov test indicates no evidence that they are drawn from different distributions ($p \approx 0.5$).
Therefore, we find no apparent relation between the periodicity values and age from this sample. 
It is worth noting that \cite{McGinnis2015} performed an analysis on a sample of dipper stars from NGC~2264, which were observed during two different epochs separated by three years. They showed that the light curve morphology of dipper stars can change from periodic to aperiodic within timescales as short as three years. Given that stars move between periodic and aperiodic accretion regimes on timescales much shorter than their PMS evolution, we do not expect to observe an appreciable difference in the age distribution between aperiodic and periodic systems, as shown in Figure~\ref{fig:MedAge}.

\begin{figure*}[!tbh]
    \includegraphics[width=0.95\textwidth]{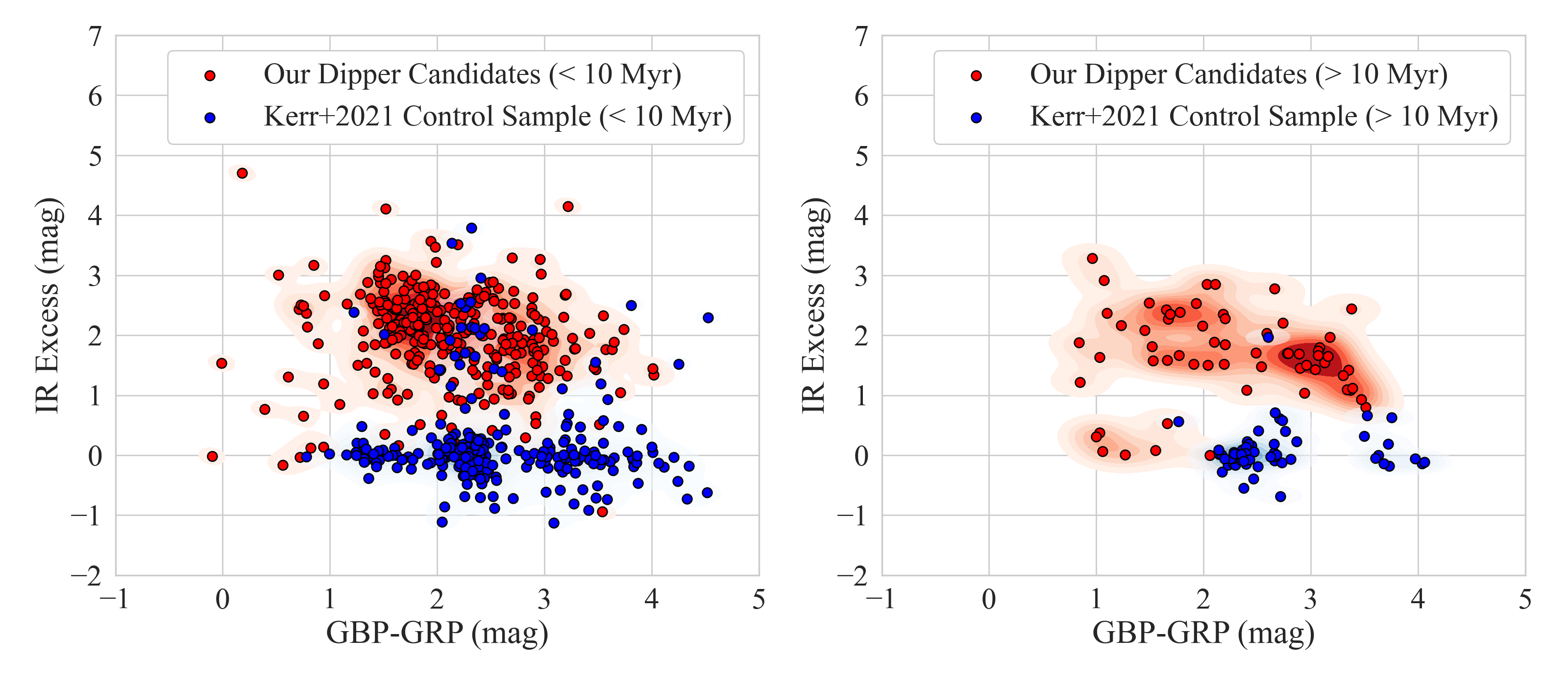}
    \caption{IR excess as a function of optical color ($GBP-GRP$) for our dipper candidates (red points).
    The \cite{Kerr2021} control sample is provided for a comparison of similarly aged sources that do not display dipper-like variability (blue points).
    Sources are split into two groups: $<10$\,Myr (left) and $>10$\,Myr (right).
    Regardless of the age division shown, we see that across all $GBP-GRP$ values, our dipper candidates exhibit greater IR excess than the K21 control sample. 
    Further, when divided by age, the distribution of our dipper candidates in this parameter space does not change appreciably.
    This suggests that the older stars in our sample may retain their disks despite being older than the nominal disk dispersal timescale of $\sim$1-3\,Myr \citep{Mamajek2009}. 
    }
    \label{fig:IRB}
\end{figure*}

We also investigated if the IR excess was more significant among the younger dipper candidates in our sample.
As shown in Figure \ref{fig:IRB}, no such evidence for a correlation between enhanced IR excess and age was found. 
In the left panel, we illustrate optical color ($GBP-GRP$) versus IR excess for dipper candidates with ages $<10$\,Myr, while the right panel illustrates this same parameter space for the sources with ages $>10 $\,Myr. 
In both panels, we illustrate the K21 control set (described in Section~\ref{subsec:ir}) for comparison. We note no significant change in IR excess as a function of age among our dipper candidate sample. 
This indicates that, regardless of age, our dipper candidates tend to be found in disk-hosting environments.
\section{Summary and Discussion}\label{sec:Discussion}
In this work, we have produced a catalog of \totcat{} young stellar variables. The catalog includes \totdippers{} dipper stars, including \newdippers{} new detections, representing the largest sample of dipper stars to date (a nearly 50\% increase in the number of new dippers in the literature).
We would like to note that nine of our new detections in the Orion Nebula Cluster were simultaneous, independent discoveries with research led by Moulton et al.~(\textit{submitted}).

Our new catalog sources have not been included in previously published dipper catalogs \citep[e.g.,][]{Morales2011,Cody_2014,Rodriguez2017,Rebull2020,tajiri2020,Nardiello2020,moulton2020search, Roggero2021,Venuti2021}.
A larger sample like ours benefits researchers in characterizing variable stars, discovering and confirming the physical mechanisms driving their variability, and investigating correlations with light curve morphology and other stellar properties. 
By creating this catalog, we hope to assist future census investigations of dipper stars.

We collected ages and associations for our catalog sources, confirming their youth. 
The variable sources in our sample exhibit a peak in the age distribution at 2.3\,Myr and 2.9\,Myr in the age estimates from the \texttt{Sagitta} neural network \citep{mcbride2020} and from \cite{Kerr2021}, respectively. 

While the light curve morphology statistics of variable stars have been relatively well studied, correlations between these characteristics and stellar age have not. We observed that the burster light curves in our catalog were only found in younger systems, while our dipper sources encompass the entire population age range of 1\,Myr to 40\,Myr.
Given that we only identified 20 burster light curves ($\approx5$\% of our catalog light curves), a larger sample of bursting light curves is required to provide more definitive evidence of such a correlation.
Similarly, we find no correlations between either of the light curve morphology metrics and stellar effective temperature.

Our investigation provides evidence that dipper stars are not purely confined to systems younger than the typical disk dispersal timescale of $\sim$1-3\,Myr \citep{Mamajek2009} (Fig~\ref{fig:AgeComp}). Moreover, we find that the older ($> 10$\,Myr) dipper candidates also exhibit IR excess akin to that of the younger ($<10$\,Myr) dipper candidates (Fig~\ref{fig:IRB}), suggesting that the older stars in our sample have retained their disks despite their age.

Recent literature concerning statistical modeling of asymmetrical events in time series data has identified potential avenues for future work in identifying and classifying dipper stars \cite{Hillenbrand2022}. 
Methods identified from such studies could improve measurements of flux asymmetry, $M$, as defined in \cite{Cody2010}. 
For example, \cite{carcea2015} and \cite{Shelef2019}
employ auto correlation functions based on Gini extremes of distributions, which would offer a more sophisticated approach to the mean function of the $M$ value. Along with these methods, autoregressive moving average (ARMA) models proposed in \cite{Trindade2010}
could provide a quantitative method for identifying and classifying dippers if they can be successfully applied to TESS light curves.

Since we finalized our target selection, TESS has continued to observe young stellar associations, including $\epsilon$~Chamaeleontis (1-3\,Myr) \citep{Feigelson2003}, the Taurus-Auriga star-forming region (${\sim}3$\,Myr) \citep[e.g.,][]{Kraus2009,Luhman2018}, the Perseus OB2 association (6\,Myr) \citep{dezeeuw1999}, and the Pleiades (125\,Myr) \citep{Stauffer1998}. 
Additionally, observations are planned in 2022 for Cepheus Flare (11\,Myr) \citep{Kerr2021}, Cerberus (30\,Myr) \citep{Kerr2021}, and Aquila (youngest stars: 1-2\,Myr) \citep{Rice2006}. 
With the continued production of high-quality light curves by the SPOC, PATHOS, QLP, and CDIPS teams, future discoveries of dipper targets are expected to expand upon this work.

\acknowledgments
We thank Ann Marie Cody, Ronan Kerr, Adam Kraus, and Tyler Moulton for their helpful conversations.
MSF gratefully acknowledges support provided by NASA through Hubble Fellowship grant HST-HF2-51493.001-A awarded by the Space Telescope Science Institute, which is operated by the Association of Universities for Research in Astronomy, In., for NASA, under the contract NAS 5-26555. 
The research has made use of the SIMBAD and VizieR databases, operated at the CDS, Strasbourg, France, and of NASA's Astrophysics Data System Abstract Service. Resources supporting this work were provided by the NASA High-End Computing (HEC) Program through the NASA Advanced Supercomputing (NAS) Division at Ames Research Center for the production of the SPOC data products.

\facilities{TESS, Gaia DR2, Gaia EDR3, ADS}
\software{Astropy\footnote{\url{http://www.astropy.org}} \citep{astropy2018}, LcTools\footnote{\url{https://sites.google.com/a/lctools.net/lctools}} \citep{schmitt2019lctools,schmitt2021lctools}, SciPy \citep{scipy},  Pandas \citep{Pandas}, Matplotlib \citep{matplotlib}, Numpy \citep{numpy}, Lightkurve version 2.0.10 \citep{Lightkurve}, Astroquery \citep{astroquery}, Astrocut (TESScut) \citep{Astrocut}, Astrobase version 0.5.3 \citep{astrobase}, cdips-pipeline \citep{cdipspipeline}, \texttt{eleanor} \citep{Feinstein2019}.}

\clearpage
\pagebreak
\bibliographystyle{aasjournalmod}
\bibliography{bibliography.bib}

\pagebreak
\begin{appendix}
\label{sec:app}

\begin{table}[h]
\caption{Variable Star Catalog}
\centering
\begin{tabularx}{\textwidth}{cccccccccccc}
{}&(1) & (2) &(3) &(4) & (5) &  (6) &   (7)&        (8)  &(9)&(10)&(11) \\
{}&        TIC &          Gaia DR2 ID &        RA &      DEC &  Tmag &  Gaia mag &    GBP &   GRP &   Vmag &     Teff &    Lum\\
{}& &  & (deg) & (deg) & (mag) & (mag) &  (mag) & (mag) & (mag) & (K) & ($\mathrm{L_{\odot}}$)\\
\specialrule{.025em}{0.25em}{0em} 
\specialrule{.025em}{0.25em}{0.05em}
0  &  355821272 &   445235933612724864 &   54.5043 &  55.1709 &   7.2 &      7.56 &   7.87 &  7.12 &   7.66 &   9966 &    NaN \\
1  &  240660823 &   391273208592115712 &    9.0644 &  48.5558 &   7.6 &      7.48 &   7.45 &  7.55 &   7.52 &  14580 &    NaN \\
2  &  390896668 &   487546103123764736 &   52.1633 &  62.4930 &   7.6 &      8.09 &   8.48 &  7.54 &   8.31 &   8914 &    NaN \\
3  &  347007416 &   420875841188703360 &    0.8631 &  55.5509 &   8.0 &      7.89 &   7.89 &  7.90 &   7.94 &      NaN &    NaN \\ 
4  &  470475885 &   272403735203771264 &   66.3523 &  53.4153 &   8.7 &      9.09 &   9.48 &  8.54 &   9.33 &      NaN &    NaN \\ 
5  &   51288359 &  5937028989511111680 &  251.9642 & -51.7678 &   8.8 &      8.91 &   9.01 &  8.62 &   9.06 &  14955 &    NaN \\ 
6  &  406956018 &   429411727913975552 &    1.5058 &  60.8672 &   9.0 &      9.29 &   9.57 &  8.85 &   9.53 &   9747 &    NaN \\ 
7  &  201357489 &   223935338503901952 &   57.4014 &  38.9821 &   9.1 &      9.46 &   9.64 &  8.93 &   9.67 &   8665 &  87.99 \\ 
8&  376088043 &   540216042984524800 &    3.4191 &  77.0364 &   9.1 &      9.58 &   9.86 &  8.97 &   9.77 &   5718 &    NaN \\ 
9  &  159089190 &  3315807764324564352 &   88.5125 &   1.6728 &   9.1 &      9.84 &  10.29 &  8.87 &  10.11 &      NaN &    NaN \\ 
\end{tabularx}

\begin{tabularx}{\textwidth}{ccccccccccccc}
\toprule
{} &   (12)  &  (13) &  (14) & (15) &(16) &  (17) &  (18) &  (19) &  (20) &(21) \\
{} &  Stellar &  Logg & LC  &  Sector &     M &     Q & Variability & Known &  Ages & Ages\\
    &  Radius  &  & Source  &    &&  && Dipper& Kerr+2021 & Our Work  \\
 {} &  ($\mathrm{R_{\odot}}$) &  (dex) & & &   &   &  & &(Myr) &  (Myr) \\
\specialrule{.025em}{0.25em}{0em} 
\specialrule{.025em}{0.25em}{0.05em} 
 0  & 19.20 & NaN & SPOC & 19 & 0.15 &  0.90 & AS & NaN & NaN & 6.1 \\ 
 1  & NaN & NaN & QLP & 17 & 0.10 &  0.81 & AS & NaN & NaN & 2.4\\ 
 2  & 22.89 & NaN & SPOC & 18 & -0.01 &  0.98 & AS & NaN & NaN & 7.5 \\ 
 3  & NaN & NaN & QLP & 17 &  0.35 &  0.92 & AD & N & NaN & 3.4\\ 
 4  & NaN & NaN & SPOC & 19 & 0.13 &  0.85 & AS & NaN & NaN & 8.5\\ 
 5  & NaN & NaN & SPOC & 12 &  0.73 &  0.61 & AD & N & NaN & 7.4\\ 
 6  & 30.41 & NaN & QLP & 18 &  0.57 &  0.60 & AD & N & NaN & 8.1\\ 
 7  & 4.16 & 3.54 & QLP & 18 &  0.30 &  0.42 &QD & N & NaN & 3.9\\ 
 8  & NaN & NaN & QLP & 18 &  0.44 &  0.63 &AD & N & NaN & 4.4\\ 
 9  & NaN & NaN & QLP & 6 &  0.75 &  0.61 & AD & N & NaN & 3.2\\ 
\toprule
\end{tabularx}
\label{tab:1}
\end{table}

Table~\ref{tab:1} displays the first ten entries of our stellar variable catalog, which includes \newdippers{} new dipper stars. The catalog is ordered by \Tmag{} and the columns are described as follows: \\
\begin{enumerate}
\item \textit{TIC ID} --- TESS source identifier
\item \textit{Gaia DR2 ID} --- Gaia Data Release 2 source identifier
\item \textit{RA} --- Right Ascension
\item \textit{DEC} --- Declination
\item \textit{Tmag} --- TESS magnitude
\item \textit{Gaia mag} --- Gaia DR2 magnitude
\item \textit{GBP} --- Gaia BP magnitude
\item \textit{GRP} --- Gaia RP magnitude
\item \textit{Vmag} --- V magnitude from TIC data
\item \textit{Teff} --- Effective temperature (K)
\item \textit{Lum} --- Luminosity ($\mathrm{L_{\odot}}$)
\item \textit{Stellar Radius} --- Stellar radius ($\mathrm{R_{\odot}}$)
\item \textit{Logg} --- Log of surface gravity (cm/s$^2$)
\item \textit{LC Source} --- Source of the pre-generated light curve
\item \textit{Sector} --- TESS observation sector
\item \textit{M} --- Flux asymmetry ``$M$" value
\item \textit{Q} --- Periodicity ``$Q$" value
\item \textit{Variability} --- The variability classification of the star based on the M and Q light curve morphology values \begin{itemize}
    \item[] For periodicity: A=Aperiodic, Q=Quasi-Periodic, P=Periodic 
    \item[] For flux asymmetry: D=Dipping, S=Symmetric, B=Bursting
\end{itemize}
\item \textit{Known Dipper} --- Indicates whether the star was classified as a dipper variable in previous investigations. Sources that are not classified as dippers (those exhibiting predominately bursting or symmetric variability) are exempt from this classification parameter.
\begin{itemize}
    \item[] Y = previously detected dipper
    \item[] N = not previously detected (new detection in this study)
    \item[] NaN = source is not a dipper variable

\end{itemize}
\item \textit{Ages Kerr+2021 (Myr)} --- Stellar age in Myr for the variable star, as determined by \cite{Kerr2021}
\item \textit{Ages Our Work (Myr)} --- Stellar age in Myr for the variable star, as determined by the \texttt{Sagitta} neural network \citep{mcbride2020}

\end{enumerate}
\end{appendix}
\end{document}